\documentclass{ws-ijtaf_none}
\usepackage{graphics}
\usepackage{epsfig}
\usepackage{amssymb}
\usepackage{natbib}

\newcommand{\vr}{\tilde{r}^{i}_{h}}
\newcommand{\vs}{\tilde{\sigma}^{i}_{h}}
\newcommand{\PandL}{{\em PandL }}

\begin{document}

\markboth{M. Bartolozzi and C. Mellen} {Local Risk Decomposition
for High-frequency Trading Systems}

\catchline{}{}{}{}{}

\title{Local Risk Decomposition for High-frequency Trading Systems}

\author{M. Bartolozzi}

\address{ Research Group, Boronia Capital, Sydney NSW
2065, Australia\\
Special Research Centre for the Subatomic Structure of Matter
(CSSM), University of Adelaide, Adelaide SA 5005, Australia \\
\email{mbartolo@physics.adelaide.edu.au} }

\author{C. Mellen}
\address{ Research Group, Boronia Capital, Sydney NSW
2065, Australia\\
\email{chris.mellen@boroniacapital.com.au} }

\maketitle

\begin{history}
\received{(Day Month Year)}
\revised{(Day Month Year)}
\end{history}

\begin{abstract}
In the present work we address the problem of evaluating the
historical performance of a trading strategy or a certain
portfolio of assets. Common indicators such as the Sharpe ratio
and the risk adjusted return have significant drawbacks. In
particular, they are global indices, that is they do not preserve
any {\em local} information about the performance dynamics either
in time or for a particular investment horizon. This information
could be fundamental for practitioners as the past performance can
be affected by the non-stationarity of financial market. In order
to highlight this feature, we introduce the {\em local risk
decomposition} (LRD) formalism, where dynamical information about
a strategy's performance is retained. This framework, motivated by
the multi-scaling techniques used in complex system theory, is
particularly suitable for high-frequency trading systems and can
be applied into problems of strategy optimization.
\end{abstract}


\section{Introduction}
\label{sec::Introduction}

Measuring the past performance of a trading system or a portfolio
of assets is one of the most important issues for financial
practitioners and portfolio managers. Evaluating performances
heavily depends on estimating ``risk"\footnote{The definition of
``risk" can be subjective and,  in fact, it does not exist a
generally accepted definition. It is often associated with the
fluctuation of returns around their mean value and thus to their
standard deviation. However, fluctuation towards positive returns
may not be considered  a form of risk. Therefore, {\em one sided}
definitions of standard deviation are also used by practitioners.
For general references on the subject the reader is referred
to~\citep{Dacorogna01,Bailey05,Meucci07}.}. In the past different
measures has been proposed but there is no general agreement about
which one is the most robust estimator for the ``quality" of a
trading strategy~\citep{Dacorogna01}.

In this paper, we contribute to the risk-adjusted performance
measurement subject by introducing a two dimensional decomposition
of the {\em profit and loss} series, {\em PandL}, of a trading
strategy. Based on this decomposition we can define a set of {\em
local} performance indicators, where ``local" refers to both time
and investment horizon. Global indicators  are then obtained via
the convolution of the decomposed signal with user-specified
kernels. The choice of the kernels, as well as their parameters,
can highlight specific features of the trading dynamics.

The relevance of this multi-scale framework,
originally developed in physics for the study of complex
system~\citep{Bouchaud99,Sornette04,Voit05},
in the contest of  risk-adjusted measures is justified by the possible
non-stationarity of the trading performances. In fact, it is
a well known fact that some strategies work well just under some
specific market condition or for a limited period of time when the
related arbitrage inefficiency has not extensively exploited yet.

 The issue of
stability in performance metrics when facing non-stationary returns
has been also addressed in econometric literature, with particular
emphasis on the {\em Sharpe ratio}~\citep{Sharpe94},
 where different nonparametric methods have been proposed in order
 to give  more ``stable" estimates,
 see \citep{Dowd00, Mukherjee04,Woehrmann05, Ledoit08} for example.
Our approach differs from the formers in many respects. Firstly,
we do not introduce a new specific risk-adjusted measure but rather a
framework where to apply the already existing ones.
Secondly, the risk associated with a strategy is not only considered as time dependent but
also as ``scale" dependent.
 Lastly, the fluctuations related to the risk
performance are estimated around {\em local trends}
in order to remove any possible bias due to some particular market condition
during the period under consideration.

The paper is structured as following: in the next section we
briefly introduce some standard indicators and point out their
drawbacks. In Sec.~\ref{sec::MRM} we introduce our {\em local risk
decomposition}, LRD, while in Sec.~\ref{sec::results} we apply the
method to the performance of different trading systems  and  we
highlight the advantages of using the LRD method if compared to
standard indicators. Discussions and conclusions are left for the
last section.

\section{Risk performance measures}
\label{sec::performance_measures}

The performance of a trading strategy are characterized by two key
quantities: the {\em cumulative return} over time, represented by
the \PandL  time series, and the {\em risk} incurred in using it.
While it is intuitive to associate profitability with the goodness
of a trading strategy, high profits can be due to lucky trades or
temporary favorable market conditions. This is the reason why
investors tend to monitor the performance of their trading systems
in time in order to recognize a possible deterioration in their
strategy.
%
 The risk-adjusted performance measures proposed in literature, see for
 example \cite{Dacorogna01}, attempt to assert the quality of a trading system by
assuming that  an investor will make his/her decision based not
only on the past returns but also on their fluctuations. Clearly
the ``amplitude" of fluctuations that a trader can tolerate
depends on his/her personal appetite for risk and is thus
subjective.
However, investors tend to be risk adverse and, in practice, a
trading strategy in order to be ``acceptable" will have to display
not only a good annualized profit but also a smooth cumulative
return or {\em PandL}. In other words, the risk related to the
fluctuation around the average return has to be small.

One of the most popular risk performance measures used in finance
is the {\em Sharpe ratio}~\citep{Sharpe94}, defined as
\begin{equation}
S=A \frac{\langle r \rangle}{\sigma_{r}}, \label{eq::sharpe}
\end{equation}
where $\langle r \rangle$ is the average return and $\sigma_{r}$
is its standard deviation. The annualization factor, $A$, is
$\sqrt{252}$ for daily returns or $\sqrt{12}$ for monthly. The
Sharpe ratio, despite being widely used, has two notable
drawbacks~\citep{Dacorogna01} among which

\begin{enumerate}
    \item It is numerically unstable for small values of $\sigma_{r}$,
    \item It does not reveal any information about the dynamics of
    the returns.
\end{enumerate}

The last point is of central interest in the present work. In
fact, since the high-frequency dynamics of the stock market is not
stationary in
time~\citep{Bartolozzi06,Bartolozzi07,Bartolozzi07b}, the
performance of trading systems can be subjected to similar
trends\footnote{Frequently, trading strategies outperform some
benchmark during a period of time by exploiting temporary
inefficiencies. Once these inefficiencies are dissipated the
performances of a trading strategy tend to deteriorate along.}.

Another widely used performance measure is the {\em risk adjusted
return}, defined as
\begin{equation}
R_{\beta}= \langle r \rangle - \beta \, \sigma_{r}.
\label{eq::risk_adjusted}
\end{equation}
This indicator, derived from  {\em utility
theory}~\citep{Dacorogna01,Bailey05}, is not affected by numerical
singularities. However, it depends on the subjective risk strength
factor, $\beta$. Furthermore, along with the Sharpe ratio, it does
not reveal any information about the evolution of the {\em PandL}.

In the next section we introduce a multi-scale framework for
estimating a risk-adjusted performance measure based on recent
work in complex system theory~\citep{Sornette04}. This framework,
while employing elementary block measures similar to
Eqs.~(\ref{eq::sharpe}) and (\ref{eq::risk_adjusted}), also
retains time and horizon information which can be fundamental in a
 strategy selection problem.

\section{The Local Risk Decomposition}
\label{sec::MRM}

In order to tackle the problem of non-stationarity of the
performance of market strategies, we introduce the {\em Local Risk
Decomposition} (LRD). The underlying idea of this method is to
extrapolate a risk measure based on the {\em local} fluctuations
of the {\em PandL}, both in time and scale (or investment
horizon). This concept is similar to the {\em detrended
fluctuation analysis}, recently proposed
 to extract correlations from non-stationary time series in the context of DNA nucleotides
sequences~\citep{Peng93}, and successively applied in finance by
several
authors~\citep{Cizeau97,Liu97,Vandewalle97,Liu99,Janosi99,Gopikrishnan00,Gopikrishnan01,Muniandy01,Matia02,Costa03,Grech04,Ivanov04,Eisler06,Bartolozzi07}.

The LRD method works as follows:

\begin{enumerate}

\item {The \PandL time series, which for high-frequency trading we
can reasonably assume to be daily updated\footnote{Note that in
high-frequency trading there is no reason for the \PandL not to be
updated intra-day or in a {\em per-trade} basis.}, $x(k)$ where
$k=1,...,N$, is divided into $M=N/h$ non-overlapping boxes of
equal sample size $h$, corresponding to different investment horizons.
In our notation $x^{i}_{h}(\bar{k})$, given $i = 1,...,M$ and $\bar{k}=(i-1)h+1,...,ih$, represents the \PandL of the
strategy under consideration  associated with the
$i^{th}$ box of length $h$.}

\item{For each box, first we perform a linear fit (that is, we
look for the local trend) of the  {\em PandL}, $y^{i}_{h}(k)$, as
well as the fluctuations around it,
\begin{equation}
\vs=\sqrt{\frac{1}{h} \sum_{\bar{k} = (i-1)h+1}^{ih}(x^{i}_{h}(\bar{k})-y^{i}_{h}(\bar{k}))^{2}}.
\end{equation}
 which we take as the {\em local risk}. The difference between the first and last point of the
 fit represents the {\em local return}, $\vr = y^{i}_{h}(ih) -
y^{i}_{h}((i-1)h+1)$, at scale $h$ for the $i^{th}$ box.

}

\item{ The procedure of points (1) and (2) is iterated over
different investment horizons $h$, in order to compare how the
trading performance changes at different scales.

}

\end{enumerate}

 It is worth noting that our
measures defined above, $\vr$ and $\vs$, are {\em local} both in
time and scale. Furthermore, the decision of taking the extremes
of the fit as a measure of the local return is  to avoid
overestimating outliers of returns that may not give a fair value
to the strategy under exam.

The next step involves the definition of the local performance
measures. In analogy with Eqs.~(\ref{eq::sharpe}) and
(\ref{eq::risk_adjusted}), we define the {\em local Sharpe ratio}
(LSR) as
\begin{equation}
\label{eq::LSR}
 S^{i}_{h}=\frac{\vr}{\vs},
\end{equation}
 and the {\em local risk adjusted return} (LRA) as
\begin{equation}
\label{eq::LRA}
 R^{i}_{h}=\vr-\beta \phi_{h} \vs,
\end{equation}
 where $\beta$  is the risk aversion of the trader (equivalent
to the $\beta$ in Eq.~(\ref{eq::risk_adjusted})) and $\phi_{h}$ is
a scaling factor, defined as
\begin{equation}
 \phi_{h} = \frac{\langle \vr \rangle_{M}}{\langle \vs
\rangle_{M}}.
\end{equation}
Now we have two dimensional representations of performance
measures that are localized {\em both} in time and investment
horizon. It is important to underline at this stage that despite
their similarities, the measures proposed in Eqs. (\ref{eq::LSR})
and (\ref{eq::LRA}) are not equivalent to those in Eqs.
(\ref{eq::sharpe}) and (\ref{eq::risk_adjusted}).

 In the next section we apply our LRD  to \PandL curves generated by different
trading strategies.

\section{Local Risk Decomposition in trading systems: applications}
\label{sec::results}

\begin{figure}
\vspace{1cm}
\centerline{\epsfig{figure=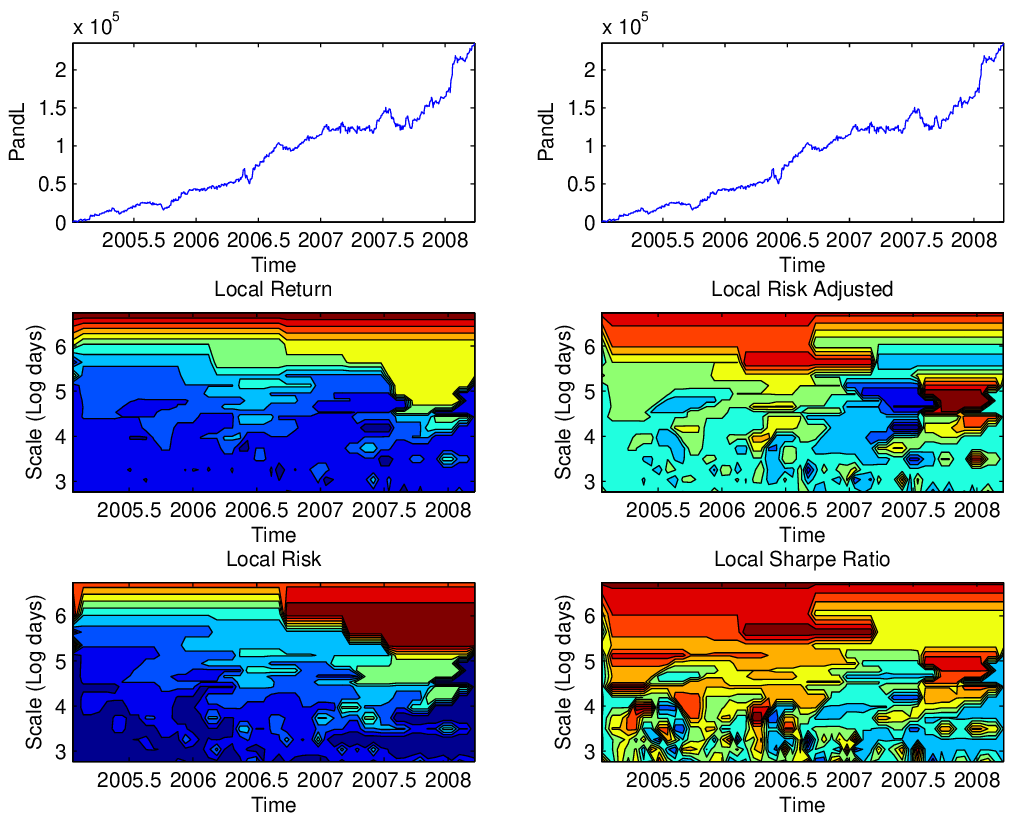,height=8cm,width=12cm}}
\caption{The two top plots are the same daily \PandL generated by
a certain trading strategy. The time series is shown twice in
order to ease the comparison with the LRD contour plots reported
underneath. On the right-hand side, we report the LRA
(middle-right, $\beta=0.75$) and the LSR (bottom-right). Both
representations capture the ``bumps" observed in 2007 and 2008. On
the left-hand side, for completeness, we show the {\em local
return}, $\vr$, (middle) and the {\em local risk}, $\vs$,
(bottom). The contour scale goes from the dark colors for the
minima to the light ones for the maxima.} \label{fig::test1}
\end{figure}

\begin{figure}
\vspace{1cm}
\centerline{\epsfig{figure=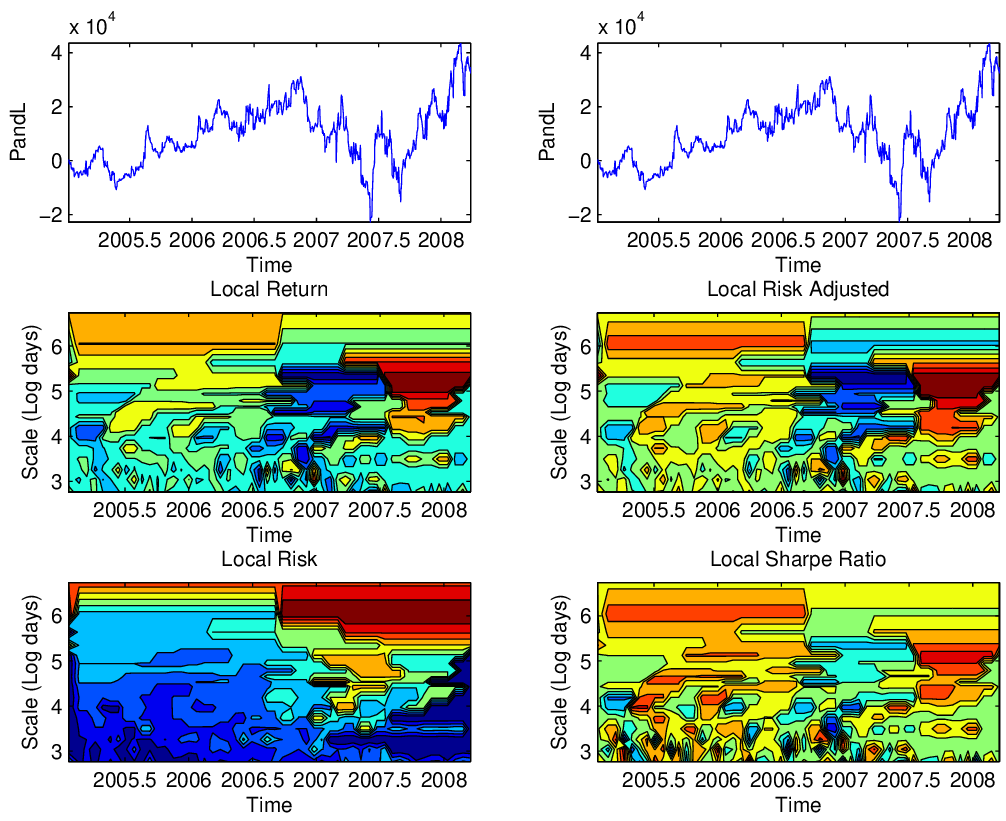,height=8cm,width=12cm}}
\caption{The plots are equivalent to those in
Fig.~\ref{fig::test1} but for a different trading system. The
performance, in this case, starts being relatively volatile from
the middle of 2006. This change in dynamics is encoded, with
different emphasis, by the LRA ($\beta=0.75$) and the LSR
measures.} \label{fig::test2}
\end{figure}

Now we consider two examples of the LRD when applied to \PandL
time series generated by different strategies. In particular, the
first time series, Fig.~\ref{fig::test1} (top), shows relatively
stationary performance over the period under consideration, with
the exception of two ``bumps" in the middle of 2007 and at the
beginning of 2008. These ``bumps" are highlighted as a valley and
a peak in the LRD, as it can be seen in the contour plots for the
LRA (Fig.~\ref{fig::test1}, middle-right, $\beta=0.75$) and for
the LSR (Fig.~\ref{fig::test1}, bottom-right).
The second time series, instead, Fig.~\ref{fig::test2} (top), is
more volatile if compared to the first: we have good performances
up to the end of 2006 when suddenly the system starts to lose
money. However, at the end of 2007 a comeback is observed. Both
LRA, (Fig.~\ref{fig::test2}, middle-right, $\beta=0.75$),  and
LSR, (Fig.~\ref{fig::test2}, bottom-right), capture this dynamics
very faithfully:  a deep valley followed by an high peak can be
observed in the last part of the time series.
The LRD framework, therefore, allows the practitioner to identify
and stress easily specific periods in time as well as specific
investment horizons that have been particularly significant during
the life (or testing) of a trading system.

\begin{figure}
\vspace{1cm}
\centerline{\epsfig{figure=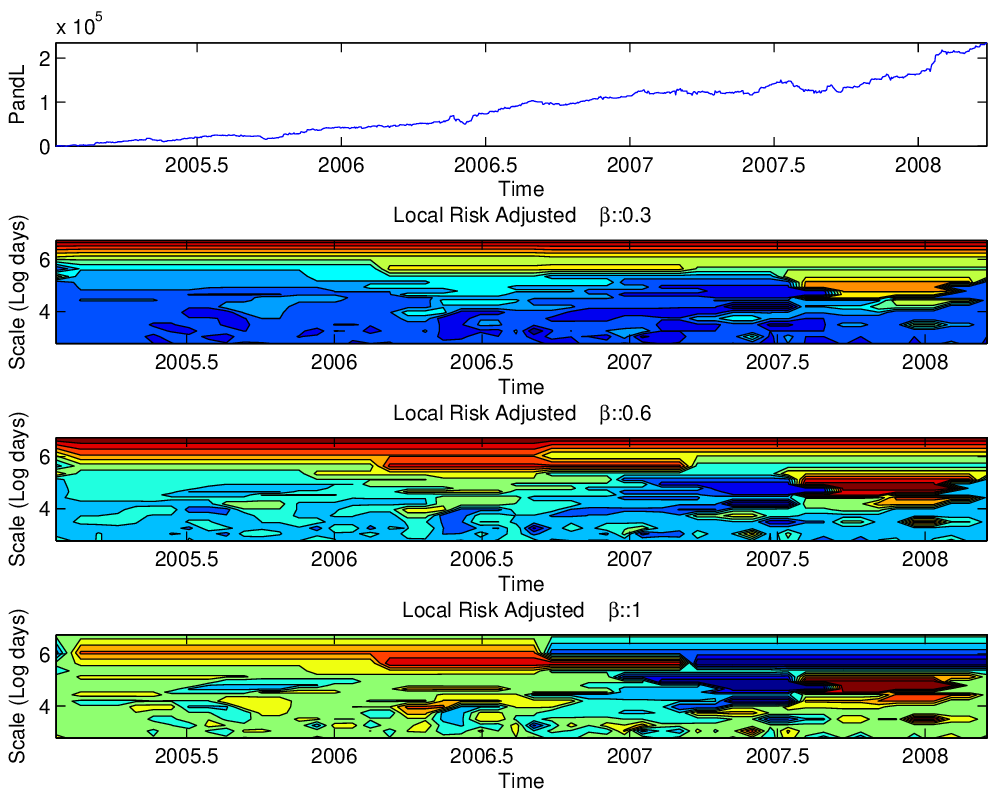,height=7cm,width=6cm}}
\caption{Different contour plots of the LRA related to the \PandL
time series of Fig.~\ref{fig::test1} (top). The different values
of $\beta$ (0.3, 0.6 and 1 from top to bottom) smooth or emphasize
volatile periods according to the different appetite for risk
chosen by the investor.} \label{fig::test3}
\end{figure}

It is important to notice that LRA and LSR magnify differently the
features of the time series presented in the former examples. This
fact is due related to the investor's particular appetite for
risk, parameterized by $\beta$ and fixed to 0.75 in
Figs.~\ref{fig::test1} and \ref{fig::test2}, that appears in the
LRA. An aggressive trader would give more importance to the
returns than to their fluctuations and, therefore, $\beta \approx
0$. By contrast, a risk adverse trader highlights the
fluctuations, so to have $\beta \approx 1$. Examples of the LRA
response to different sensitivities are shown in
Fig.~\ref{fig::test3} for the first time series.

%

%

\section{Extracting performance indices from the LRD}
\label{sec::performance_indicator}

In the previous section, we introduced a framework to estimate
local risk measures from the \PandL of a trading strategy. The
complete time/scale decomposition, despite being a faithful
representation of the {\em PandL}'s dynamics, as well as visually
appealing, can be cumbersome to use in practical applications,
such as algorithms for strategies optimization. It is, therefore,
of interest to derive a single performance indicator from the
information provided by the LRD.

The advantage of having a LRD of the \PandL signal lies in the
possibility to ``customize" the final indicator according to the
user's specific need. In other words, different traders may focus
on different investment horizons or could be more interested in
limited periods of time characterized by specific market
condition: these preferences can be encoded in the integration of
the LRD.
 In fact, for a generic local performance measure, $f$, (LRA or LSR, for
example) we define our indicator as the convolution of this
quantity with time/scale kernels over the ranges of interest as, for instance, from $T_{0}$ to $T$ for
time and from $h_{min}$ to $h_{max}$ for the investment horizon. In order to ease the notation,
we assume a continuous decomposition for the \PandL , that is $i
\rightarrow t$ and $h \rightarrow s$, and we define an LRD indicator as
\begin{equation}
 \Phi^{f}_{\tau,\rho} = \frac{\int_{h_{min}}^{h_{max}} ds \, K_{s}(\frac{s-\rho}{\delta s}) \, \eta_{f}(\tau,s)}{ \int_{h_{min}}^{h_{max}} ds \,
K_{s}(\frac{s-\rho}{\delta s})},
 \label{eq::indicator}
\end{equation}
where
\begin{equation}
\eta_{f}(\tau,s)=\frac{\int_{T_{0}}^{T} dt \, K_{t}(\frac{t-\tau}{\delta t})
\, f(t,s)}{\int_{T_{0}}^{T} dt \, K_{t}(\frac{t-\tau}{\delta t})},
\label{eq::eta}
\end{equation}

being $K_{s}$ and $K_{t}$ convolution kernels,  $\rho$ and $\tau$
representing the ``principal" investment horizon and time while
$\delta t $ and $\delta s$ are dilatation
coefficients~\citep{Silverman96}. These parameters can be tuned
for different investor's requirements, making the method
particularly flexible. For example, by using hard kernels such as
the Heaviside function, it is possible to cut the contribution of
the performance beyond some specified look-back period. Otherwise,
if it is preferred to give a weight to whole the historical
performance of the trading strategy, a Gaussian kernel would be
suitable.


In order to underline the flexibility of our indicator
$\Phi^{f}_{\tau,\rho}$, we perform a numerical test on two
artificial \PandL time series for different choices of the
parameters in Eqs.~(\ref{eq::indicator}) and (\ref{eq::eta}). We
restrict the range of choices by fixing $\tau=\max(t)$, since
investors tend to give more importance on the recent performance
of their strategies. The dilatation coefficients are selected
according to: $\delta s = 100 \rho$ and $ \delta t = \left[
\max(t) - \min(t) \right]/4$. The errors on the estimates have
been calculated via the jacknife method~\citep{Kunsch89} and
indicated between brackets as uncertainty in the last digit.

\begin{figure}
\vspace{1cm}
\centerline{\epsfig{figure=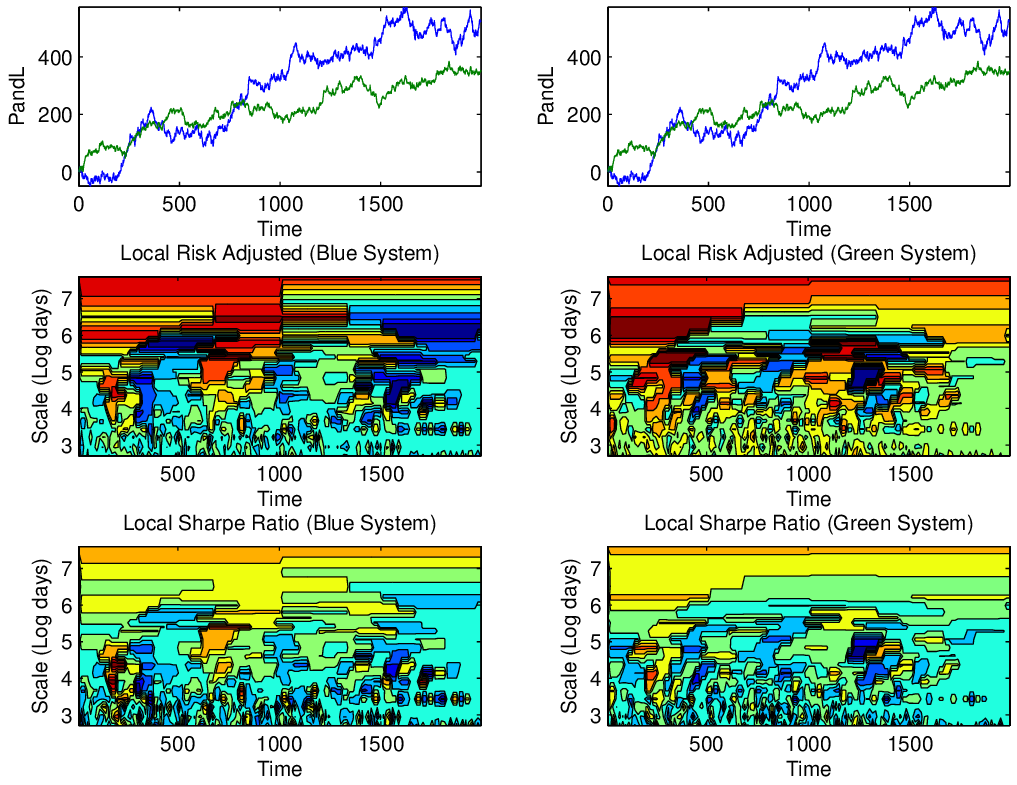,height=8cm,width=12cm}}
\caption{LRD for two simulated \PandL. The left hand side
corresponds to the blue time series and the right hand side to the
green. The noise amplitude for the blue time series is 1.5 times
the green one. The Sharpe ratio, calculated via
Eq.~(\ref{eq::sharpe}) is 0.7(2) for both time series and,
therefore, they are undistinguishable according to this
performance indicator. For the LRA we used $\beta=0.75$.}
\label{fig::pandl-compare-syn}
\end{figure}

The LRD of two artificially generated {\em PandL}, each with 2000
data points, with different linear drifts as well as a different
superimposed noise amplitude, is shown in
Fig.~\ref{fig::pandl-compare-syn}. The first time series (blue)
provides a better return at the expenses of higher volatility. The
second time series (green), in contrast, exhibits a relatively
stable growth. Despite the intrinsic differences, the annualized
Sharpe ratio, Eq.(\ref{eq::sharpe}),
 results to be the same for the two time series, namely $S=0.7(2)$,
  making them look equivalent from its prospective.
On the other hand, the LRD framework gives a much broader picture
regarding the performances of the two time series. The results are
summarized in Table 1 and Table 2.
In the first one, we report for different principal investment
horizons, $\rho$, the values of $\Phi^{LRA}_{\tau,\rho}$ and
$\Phi^{LSR}_{\tau,\rho}$ for the two trading systems when a
uniform kernel is used, $K_r \equiv K_t \equiv 1$. In the second
table, instead, we show the same results for Gaussian kernels.
We also report the values of $\eta_{LRA}(\tau,s)$
and $\eta_{LSR}(\tau,s)$ at the scale of
main interest, that is for $s \equiv \rho$. The reason behind this
is that these two quantities, which are nothing but kernel
weighted averages of a risk-adjusted performance, $f$,
over time being the scale $s$ fixed, Eq.(\ref{eq::eta}), can be considered as further
performance indicator when the strategy has a characteristic time scale (holding period),
 $\rho$ in this case\footnote{In the discrete algorithm described
in Sec.~\ref{sec::MRM} the weighted average would be over the $M$ risk-adjusted measures estimated
over the boxes of length $h$.}.

%
 The results show that when we do not apply any convolution kernel, Table 1,
the performance indicators $\Phi^{LRA}_{\tau,\rho}$ and
$\Phi^{LSR}_{\tau,\rho}$ would pick the blue strategy, that is,
the one with the highest return, as the best out of the two.
However, if we consider the indicator at a specific investment
horizon $\rho$, that is $\eta_{LRA}(\tau,\rho)$ and
$\eta_{LSR}(\tau,\rho)$, the situation is not as clear. On the
other hand, when the indicators are extracted via two Gaussian
kernels centered in the last day of trade and at the horizon
$\rho$, explicitly giving more importance to a particular
time/scale region, the best performing system would be the green
one. This result is due to the fact that the blue strategy is not
performing well in the last period of the \PandL series where  the
time kernel is centered\footnote{It is also worth to notice that
 the differences between the estimates of the risk-performance measures
 discussed in this paragraph are significant based on the error estimates,
  obtained via the jacknife method, reported in the same tables.}.
This simple example highlights the flexibility of the LRD
framework when compared to global quantities such as the Sharpe
ratio which ignore the dynamics of the performance.

\begin{table}[h]
\label{tab::uniform} \tbl{The values for  $\Phi^{LRA}_{\tau,\rho}$
and $\Phi^{LSR}_{\tau,\rho}$ along with $\eta_{LRA}(\tau,\rho)$
and $\eta_{LSR}(\tau,\rho)$  for different $\rho$. The values on
the left refer to the first time series (blue) in
Fig.~\ref{fig::pandl-compare-syn}, while the values on the right
refer to the second time series (green). The kernel used for the
integration of Eq.~(\ref{eq::indicator}) is uniform with $K_r
\equiv K_t \equiv 1$. The LRA  has been normalized by its standard
deviation over the time/scale boxes. This procedure is not
fundamental for asserting the performances of a strategy. In
brackets is the error of the last digit calculated via the
jacknife method.} {\begin{tabular}{@{}cccccc@{}} \toprule &
$\rho=50$  & $\rho=100$ & $\rho=250$ & $\rho=500$ & $\rho=1000$ \\
\colrule
$\Phi^{LRA}_{\tau,\rho}$\hphantom{00} & $0.36(1)/0.34(1)$ & $0.36(1)/0.34(1)$ & $0.364(9)/0.34(1)$ & $0.36(1)/0.34(1)$ & $0.36(1)/0.34(1)$  \\
$\Phi^{LSR}_{\tau,\rho}$\hphantom{00} & $2.71(7)/2.4(1)$ & $2.71(8)/2.4(1)$ & $2.7(1)/2.37(9)$ & $2.71(6)/2.37(9)$ & $2.7(1)/2.4(9)$ \\
$\eta_{LRA}(\tau,\rho)$\hphantom{0} & $0.018(6)/0.02(1)$ & $0.040(4)/0.04(1)$ & $0.05(1)/0.097(9)$ & $0.23(1)/0.33(2)$ & $0.42(2)/0.51(2)$ \\
$\eta_{LSR}(\tau,\rho)$\hphantom{0} & $0.5(2)/0.5(2)$ & $0.64(8)/0.6(2)$ & $0.5(1)/0.9(2)$ & $1.53(9)/2.2(2)$ & $2.19(6)/2.6(1)$  \\
\botrule
\end{tabular}}
\end{table}

\begin{table}[h]
\label{tab::gauss} \tbl{Same as Table 1 but using two Gaussian
kernels in Eqs.~(\ref{eq::indicator})-(\ref{eq::eta}). For the
calculation, $\tau=\max(t)$ while $\delta s = 100 \rho$, and $
\delta t = \left[ \max(t) - \min(t) \right]/4$.}
{\begin{tabular}{@{}cccccc@{}} \toprule & $\rho=50$  & $\rho=100$
& $\rho=250$ & $\rho=500$ & $\rho=1000$ \\ \colrule
$\Phi^{LRA}_{\tau,\rho}$\hphantom{00} & $-0.12(2)/0.21(2)$ & $-0.12(2)/0.21(4)$ & $-0.11(2)/0.21(4)$ & $-0.11(1)/0.22(4)$ & $-0.11(2)/0.22(3)$  \\
$\Phi^{LSR}_{\tau,\rho}$\hphantom{00} & $1.60(6)/1.9(1)$ & $1.63(6)/1.9(2)$ & $1.64(5)/1.9(1)$ & $1.64(4)/1.9(1)$ & $1.64(8)/1.9(1)$ \\
$\eta_{LRA}(\tau,\rho)$\hphantom{0} & $0.00(2)/0.017(9)$ & $-0.12(3)/-0.03(3)$ & $0.13(7)/-0.02(7)$ & $-0.58(3)/0.2(1)$ & $0.02(5)/0.43(2)$ \\
$\eta_{LSR}(\tau,\rho)$\hphantom{0} & $0.4(1)/0.3(1)$ & $-0.1(2)/0.2(3)$ & $0.1(1)/0.6(1)$ & $0.1(1)/2.5(3)$ & $1.72(9)/2.5(1)$  \\
\botrule
\end{tabular}}
\end{table}

As for more traditional indicators, the LDR framework can be
used in the contest of investment optimization. In high-frequency trading, for example, a typical problem
could be how to distribute the capital allocation among several different independent trading strategies
 applied over the same contract\footnote{Usually high-frequency strategies have a relatively small volume capacity
for each single contact given that they rely on a market impact close to zero at any moment in time.
 One way to get around this issue and push more volume into the market is to scale up the number of trading
strategies and, effectively, to build a portfolio of them.}.
 In this case the weight vector is
determined via an optimization procedure of one or more of the
risk-adjusted measures, the target variables, estimated from the {\em PandL}
curves obtained in the backtesting.
In this contest, the use of the LRD as target variable
can be interpreted as a ``weighted" optimization
where more emphasis can be placed, for example, on the last
performance period and on a characteristic time scale associated with the
strategies in question. For some more detailed discussions on trading strategies optimization
in the high-frequency space, which goes beyond the scope of this work, the interested reader can
refer to~\citep{Chan09,Aldridge10}.

\section{Discussion and conclusions}
\label{sec::Conclusions}

In the present paper we have introduced a local risk decomposition
framework that retains dynamical information about the performance
of a trading strategy. This framework is very useful for
practitioners who work at high-frequencies as it provides a map of
the non-stationarity and multiscale features of the \PandL time
series.
Moreover, from the LRD it is possible to construct a single
indicator for the performance of the trading system, as shown in
Sec.~\ref{sec::performance_indicator}. The advantage of this
indicator when compared to more traditional ones, such as the
Sharpe ratio for example, lies in the fact that the user can
choose to put more emphasis on some period in time or some
specific investment horizons according to his/her preference. It
is also important to stress the {\em local} detrending procedure
in the risk estimate which we have used in order to take into
account for the possible non-stationarity of the time series.

On the other hand, in order to have a reliable estimation of the
dynamics at different scales, the LRD requires a reasonable amount
of samples in the {\em PandL}. This drawback makes the LRD more
suitable for high/medium frequency trading systems rather than log
term ones.

 In conclusion, the LRD framework
can be a useful alternative to more traditional risk-adjusted
performance indicators and, consequently, it can be applied to optimization problems
such as the creation of a portfolio of different high-frequency trading systems.

%

\section*{Acknowledgement}
The authors would like to thank Richard Grinham and David Fussell
for a careful reading of the manuscript.

\bibliographystyle{plain}
\bibliography{multiscaleRiskSub}

\begin{thebibliography}{10}

\bibitem{Aldridge10}
Irene Aldridge.
\newblock {\em High-frequency trading: a practical guide to algorithmic
  strategies and trading systems}.
\newblock Wiley \& Sons, New Jersey, 2010.

\bibitem{Bailey05}
Roy~E. Bailey.
\newblock {\em The economics of financial markets}.
\newblock Cambridge University Press, Cambridge, UK, 2005.

\bibitem{Bartolozzi06}
Marco Bartolozzi, Derek~B. Leinweber, and Anthony~W. Thomas.
\newblock Scale-free avalanche dynamics in the stock market.
\newblock {\em Physica A}, 370:132--139, 2006.

\bibitem{Bartolozzi07b}
Marco Bartolozzi, Christopher Mellen, Francis Chan, David Oliver, Tiziana
  Di~Matteo, and Tomaso Aste.
\newblock Applications of physical methods in high-frequency futures markets.
\newblock {\em Proc. of SPIE}, 6802:680203, 2007.

\bibitem{Bartolozzi07}
Marco Bartolozzi, Christopher Mellen, Tiziana Di~Matteo, and Tomaso Aste.
\newblock Multi-scale correlations in different futures markets.
\newblock {\em The European Physical Journal B}, 58(2):207--220, 2007.

\bibitem{Bouchaud99}
J.-P. Bouchaud and M.~Potters.
\newblock {\em Theory of financial risk}.
\newblock Cambridge University Press, Cambridge, 1999.

\bibitem{Chan09}
Ernest~P. Chan.
\newblock {\em Quantitative Trading}.
\newblock Wiley \& Sons, New Jersey, 2009.

\bibitem{Cizeau97}
Liu~Y.H. Cizeau, P.~and, M.~Meyer, K.C. Peng, and H.E. Stanley.
\newblock Volatility distribution in the s\&p stock index.
\newblock {\em Physica A}, 245:441--445, 1997.

\bibitem{Costa03}
R.L. Costa and G.L. Vasconcelos.
\newblock Long-range correlations and nonstationarity in the brazilian stock
  market.
\newblock {\em Physica A}, 329:231--248, 2003.

\bibitem{Dacorogna01}
Michael~M. Dacorogna, Ramazan Gen\c{c}ay, Ulrich M\"uller, Richard~B. Olsen,
  and Olivier~V. Pictet.
\newblock {\em An introduction to high-frequency finance}.
\newblock Academic Press, San Diego, 2001.

\bibitem{Dowd00}
Kevin Dowd.
\newblock Adjusting for risk: an improved sharpe ratio.
\newblock {\em International Review of Economics and Finance}, 9:209--222,
  2004.

\bibitem{Eisler06}
Z.~Eisler and J.~Kert$\rm\acute{e}$sz.
\newblock Liquidity and the multiscaling properties of the volume traded on the
  stock market.
\newblock {\em Europhys. Lett.}, 77:28001, 2007.

\bibitem{Gopikrishnan01}
P.~Gopikrishnan, V.~Plerou, X.~Gabaix, L.A.N. Amaral, and H.E. Stanley.
\newblock Price fluctuations and market activity.
\newblock {\em Physica A}, 299:137--143, 2001.

\bibitem{Gopikrishnan00}
P.~Gopikrishnan, V.~Plerou, Y.H. Liu, Gabaix~X. Amaral, L.A.N.~and, and H.E.
  Stanley.
\newblock Scaling and correlation in financial time series.
\newblock {\em Physica A}, 287:362--373, 2000.

\bibitem{Grech04}
D.~Grech and Z.~Mazur.
\newblock Can one make any crash prediction in finance using the local hurst
  exponent idea?
\newblock {\em Physica A}, 336:113--145, 2004.

\bibitem{Ivanov04}
Plamen~Ch. Ivanov, Ainslie Yuen, Boris Podobnik, and Youngki Lee.
\newblock Common scaling patterns in intertrade times of u.s. stocks.
\newblock {\em Phys. Rev. E}, 69:56107, 2004.

\bibitem{Janosi99}
I.M. J$\acute{\rm a}$nosi, B.~Janecsk$\acute{ \rm o}$, and I.~Kondor.
\newblock Statistical analysis of 5 s index data of the budapest stock
  exchange.
\newblock {\em Physica A}, 269:111--124, 1999.

\bibitem{Kunsch89}
H.~R. Kunsch.
\newblock The jacknife and the bootstrap for general stationary observations.
\newblock {\em The Annals of Statistics}, 17:1217--1241, 1989.

\bibitem{Ledoit08}
Oliver Ledoit and Michael Wolf.
\newblock Robust performance hypotesis testing with the sharpe ratio.
\newblock {\em Journal of Empirical Finance}, 15:850--859, 2008.

\bibitem{Liu97}
Y.H. Liu, P.~Cizeau, M.~Meyer, K.C. Peng, and H.E. Stanley.
\newblock Scaling behavior in economic time series.
\newblock {\em Physica A}, 245:437--440, 1997.

\bibitem{Liu99}
Y.H. Liu, P.~Gopikrishnan, P.~Cizeau, M.~Meyer, K.C. Peng, and H.E. Stanley.
\newblock The statistical properties of the volatility price fluctuations.
\newblock {\em Phys. Rev. E}, 60:1390--1400, 1990.

\bibitem{Matia02}
K.~Matia, L.A.N. Amaral, S.P. Goodwin, and H.E. Stanley.
\newblock Different scaling behaviors of commodity spot and future prices.
\newblock {\em Phys. Rev. E}, 66:045103(R), 2002.

\bibitem{Meucci07}
Attilio Meucci.
\newblock {\em Risk and Asset Allocation (third edition)}.
\newblock Springer, Berlin, 2007.

\bibitem{Mukherjee04}
D.~Mukherjee and A.~Ullah.
\newblock Nonparametric sharpe ratio.
\newblock {\em Journal of Quantitative Economics}, 2(2):172--185, 2004.

\bibitem{Muniandy01}
S.V. Muniandy, S.C. Lim, and R.~Murugan.
\newblock Inhomogeneous scaling behaviors in malaysian foreign currency
  exchange rates.
\newblock {\em Physica A}, 301:407--428, 2001.

\bibitem{Peng93}
C.-K. Peng, S.~V. Buldyrev, S.~Havlin, M.~Simons, H.~E. Stanley, and A.~L.
  Goldberger.
\newblock Mosaic organization of {DNA} nucleotides.
\newblock {\em Physical Review E}, 49(2):1685--1689, Feb 1994.

\bibitem{Sharpe94}
W.F. Sharpe.
\newblock The sharpe ratio.
\newblock {\em Journal of Portfolio Managment}, 21:49--59, 1994.

\bibitem{Silverman96}
B.W. Silverman.
\newblock {\em Density estimation for statistics and data analysis}.
\newblock Chapman-Hall, London, 1996.

\bibitem{Sornette04}
Didier Sornette.
\newblock {\em Critical phenomena in natural sciences}.
\newblock Springer-Verlag, Berlin, 2004.

\bibitem{Vandewalle97}
N.~Vandewalle and M.~Ausloos.
\newblock Coherent and random sequences in financial fluctuations.
\newblock {\em Physica A}, 246:454--459, 1997.

\bibitem{Voit05}
Johannes Voit.
\newblock {\em The Statistical Mechanics of Financial Markets}.
\newblock Spriger-Verlag, Berlin, 2005.

\bibitem{Woehrmann05}
Peter Woehrmann, Willy Semmler, and Martin Lettau.
\newblock Nonparametric estimation of time-varying sharpe ratio in dynamic
  asset pricing models.
\newblock {\em Working paper 225, Institute for Empirical Research in Economics
  University of Zurich}, 2005.

\end{thebibliography}

\end{document}